\documentclass[conference]{IEEEtran}
\usepackage{cite}
\usepackage{graphicx}
\usepackage{float}
\usepackage{amsmath}
\usepackage{listings}
\usepackage{xcolor}
\usepackage{subfig}
\usepackage{url}

\usepackage{tikz}
\usepackage{textcomp}
\usepackage{hyperref}
\usepackage{lipsum}

\newcommand\copyrighttext{%
  \footnotesize \textcopyright 2020 IEEE. Personal use of this material is permitted. 
    Permission from IEEE must be obtained for all other uses, in any current or future media, 
    including reprinting/republishing this material for advertising or promotional purposes, 
    creating new collective works, for resale or redistribution to servers or lists, 
    or reuse of any copyrighted component of this work in other works.
  DOI: \href{https://doi.org/10.1109/ICCCNT49239.2020.9225310}{10.1109/ICCCNT49239.2020.9225310}}
\newcommand\copyrightnotice{%
\begin{tikzpicture}[remember picture,overlay]
\node[anchor=south,yshift=10pt] at (current page.south) {\fbox{\parbox{\dimexpr\textwidth-\fboxsep-\fboxrule\relax}{\copyrighttext}}};
\end{tikzpicture}%
}

\title{Customer Analytics using Surveillance Video}
\author{ }
\date{May 2020}

\author{
\IEEEauthorblockN{Dr. Earnest Paul Ijjina\IEEEauthorrefmark{1}}
\IEEEauthorblockA{Assistant Professor, Department of Computer Science and Engineering,\\
	National Institute of Technology Warangal, India-506004\\
	Email : \IEEEauthorrefmark{1}  iep@nitw.ac.in}

\IEEEauthorblockN{Aniruddha Srinivas Joshi\IEEEauthorrefmark{2}, Goutham Kanahasabai\IEEEauthorrefmark{3}, Keerthi Priyanka\IEEEauthorrefmark{4}}
\IEEEauthorblockA{B.Tech., Department of Computer Science and Engineering,\\
	National Institute of Technology Warangal, India-506004\\
	Email : \IEEEauthorrefmark{2}aniruddha980@gmail.com, \IEEEauthorrefmark{3}gauthamkanags@gmail.com, \IEEEauthorrefmark{4}keerthi51299@gmail.com}
}

\begin{document}
\maketitle
\copyrightnotice

\begin{abstract}
The analysis of sales information, is a vital step in designing an effective marketing strategy. This work proposes a novel approach to analyse the shopping behaviour of customers to identify their purchase patterns. An extended version of the Multi-Cluster Overlapping $k$-Means Extension (MCOKE) algorithm with weighted $k$-Means algorithm is utilized to map customers to the garments of interest. The age \& gender traits of the customer; the time spent and the expressions exhibited while selecting garments for purchase, are utilized to associate a customer or a group of customers to a garments they are interested in. Such study on the customer base of a retail business, may help in inferring the products of interest of their consumers, and enable them in developing effective business strategies, thus ensuring customer satisfaction, loyalty, increased sales and profits.
\end{abstract}
\begin{IEEEkeywords}
Customer Analytics, Weighted $k$-Means, Customer Demographics
\end{IEEEkeywords}

\section{Introduction}
The analysis of customers behaviour in retail stores gives valuable and key information, from a marketing perspective. In fact, analysis of human behaviour using video surveillance systems is an emerging area of research in the field of computer vision. Understanding customer expectations is a complex problem, which could be better addressed by segmenting customers based on their demographics and behavioural cues, to help retailers figure out the exact demands of the current consumer population. This can also help them in devising effective marketing strategies, that can target the specific preferences of the consumers.

Consumer behaviour analysis is vital to the success of a business as it provides insight into the factors that influence a consumer's decision to buy a particular product of interest. Through this study, it is possible to fill the gap between customers expected products and the available products at a store, thereby improving the effective use of floor space and maximizing customer satisfaction; sales and profits of the store.

According to a report from Salesforce \cite{ref2}, 76\% of consumers expect the companies to understand their needs and expectations. This suggests that, if the retailers are unable to understand the customer's needs before the consumer explicitly state them. Otherwise, the store will lose that customer. Video surveillance of public spaces through Closed Circuit TeleVision (CCTV) cameras is used widely to monitor the location and the behavioural cues of people in those places \cite{ref1}. There has been an increase in demand for video surveillance systems to ensure the safety of citizens in public spaces and hence more CCTV cameras are being deployed in common places. In this work, we propose to leverage the existing surveillance systems to investigate the behaviour of customers in retail stores. In stores, products are displayed to attract customers and engage customers to promote more sales. The analysis of customer behaviour though video surveillance systems is needed to re-evaluate the product placement and marketing strategies, to ensure better sales in the future. Due to the aforementioned reasons, this work aims to capture the way the customers interact with products to develop a robust framework to identify customer preferences; that could in-turn be utilized in developing effective marketing strategies to maximize sales.

\section{Related Works and Literature}
The domain of customer behaviour analysis and sales analysis has become an active area of research in the past few decades due to the use of technology to capture and monitor sales. Ko \cite{ref21a} mentions that customer behaviour analysis from video analysis primarily comprised of video acquisition, object detection, object tracking, motion detection, behaviour analysis, person identification and visualisation. This study utilized background subtraction, temporal difference and optical flow to detect motion.

An approach for customer behaviour analysis was proposed by Haritaogla \textit{et al.} \cite{ref4}, in which a system is developed to identify shopping groups by detecting and tracking various consumers that are waiting at the checkout counter. The people were clustered into shopping groups by analysing inter-body distances. In Popa \textit{et al.} \cite{ref3}, a system to extract features from the tracked path (such as the relation with the shopping place, head orientation, direction and speed of walking, etc.) are used to assess the customer's appreciation (positive or negative) of the product, thereby capturing the customer behaviour. Background subtraction is utilised to detect customers at the entry points and were tracked with mean shift algorithm. The task of customer behaviour analysis involves many other complex activities beyond trajectories, such as adding a product to the shopping cart, standing near a shelf, picking up a product, glancing etc., which makes this a more challenging area of research. 

Human behaviour analysis is also expression detection, pose estimation, gaze identification etc. In Chen \textit{at al.} \cite{ref22}, a star-like representation is used to represent human pose, human actions were modelled as a sequence of these representations and a Gaussian Mixture Model (GMM) is used to recognize human actions. Abe \textit{et al.} \cite{ref23}  proposed an approach to detect head orientation of pedestrians in surveillance video using background subtraction with histograms of oriented gradients (HOG) and support vector machine (SVM) to detect moving pedestrians, followed by template matching to detect face region and an SVM to estimate the direction of the face. Juan \textit{et al.} \cite{ref25} proposed an approach to utilise the video sequence represented as collection of spatial and spatio-temporal features by extracting static and dynamic interest points. A hierarchical model combining both spatial and spatio-temporal features is used for recognising human actions.

Coming to more recent methodologies, Maulana \textit{et al.} \cite{ref26} proposed an AI-based customer behaviour analytics system. Traditionally, customer behaviour analysis, which requires object detection algorithms are normally implemented on state of the art GPUs or cloud based platforms. The authors propose a system that utilised edge devices such as embedded computers. By importing an existing deep-learning based customer behaviour analysis into such edge devices, they have minimised the cost, power and space needed to implement such systems. Kulkarni \textit{et al.} \cite{ref27} proposed a multi-agent system for analysing customer behaviour. The authors state that the path traversed by a customer is depictive of his/her behavioural pattern, preferences and needs. The overall path of a customer is used to deduce buying patterns and gain knowledge about the items that he/she plans to purchases at a future time. Such traversal paths are associated to help deduce the behaviour of similar customers, enabling them to suggest personalised paths and rearrange items in stores to maximise profits.\par
There is little work done in the field of customer analytics in the Indian context. Performing customer analytics in the Indian context is challenging due to the distinct cultural and behavioral characteristics between the Indian and western setting.\par

The reminder of the paper is organized so that Section \ref{prop} explains the proposed approach, Section \ref{study} presents the experimental results \& analysis followed by Section \ref{conc} on conclusions and future work.

\section{Proposed Approach}\label{prop}
\begin{figure} 
	\centering
	\includegraphics[width=85mm, height=85mm]{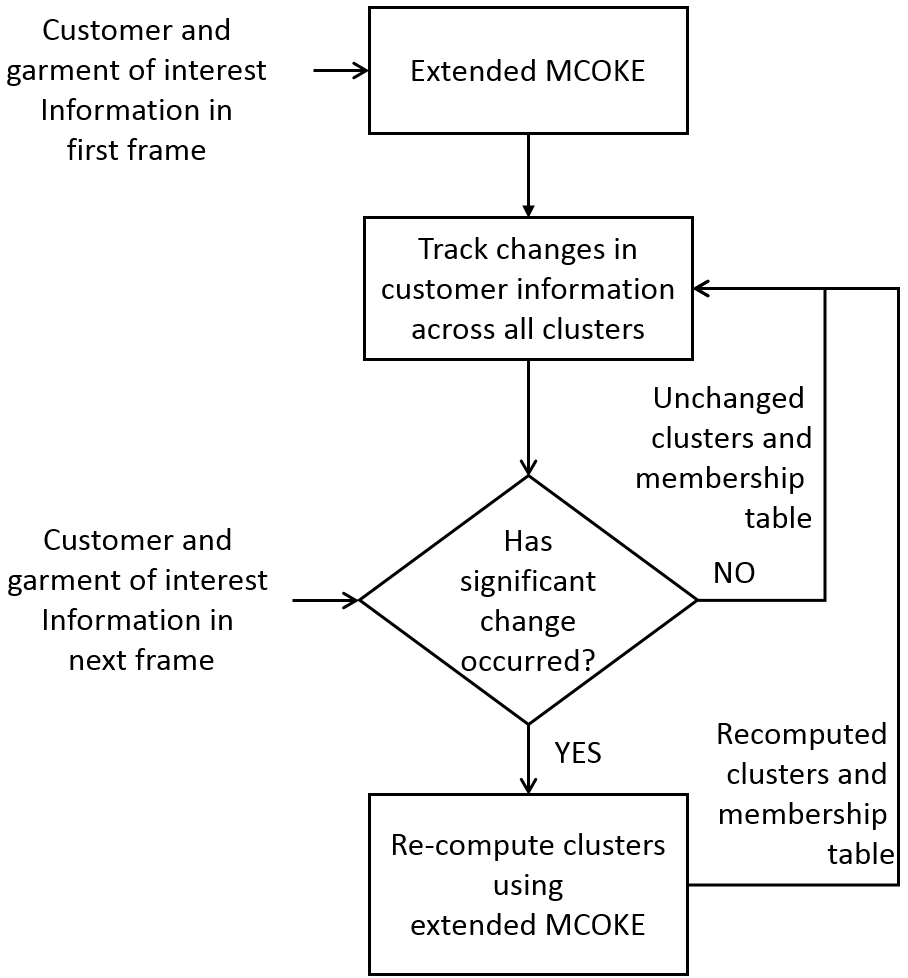}
	\caption{Workflow of the proposed Framework}
	\label{fig:workflow}
\end{figure}

This section describes the proposed framework illustrated in Figure \ref{fig:workflow}. The proposed approach aims to associate a customer or a group of customers to a garment on which they showed interest. We assume the prior-computation of (a) the gender, age, expression of the customer (tagged by a unique \textit{Tracking-ID} for each customer present in the surveillance video) along with their spatial location represented by bounding box and, (b) the colour, location (as bounding box) and \textit{Tracking-ID} of all garments of interest; for the surveillance videos in the dataset. This approach assumes that the above information pertaining to the customers and their preferred garments is already present in the dataset.

The proposed approach starts by mapping the customers in the store, to the garments of interest. Some of the possible scenarios during sales are i) a customer is interested in multiple garments, ii) a given garment is of interest to multiple customers, and iii) the set of garments a customer finds interesting may change over time as the customer shuffles through various garments. In a nutshell, there is a \textit{many-to-many} mapping between customers and garments and the mapping between a customer and garments may change over time.

To capture the mapping of a customer to his garments of interest, a modified Multi-Cluster Overlapping $k$-Means Extension (MCOKE) algorithm proposed by Said Baadel \textit{et al.}\cite{Baadel2015MCOKEMO} is used. The MCOKE algorithm is extended to use the Weighted $k$-Means algorithm (WKM) \cite{10.1007/11546849_48} in place of the $k$-Means algorithm in the initial step of MCOKE, to create clusters. The coordinates of the centre's of the bounding boxes of all garments of interest present in the beginning frame of the surveillance video are treated as the initial $k$ \textit{centroid} points, where $k$ denotes the total number of garments of interest present in the frame. These \textit{centroid} points are assigned a high weight value (\textit{10} units). The centre \textit{point}'s of the bounding boxes of the customers in the store are assigned a weight value of \textit{1} unit.

In the first step, the Weighted $k$-Means (WKM) algorithm \cite{10.1007/11546849_48} is applied to group the given \textit{point}'s into clusters, so that the \textit{point}'s present within a maximum distance (\textit{maxDist}) will become the new \textit{centroid} points. The \textit{point}'s corresponding to the centres of the garments of interest are given a high weight value compared to the centre \textit{point} of the bounding box of the customer, to ensure that the new \textit{centroid} points lie closer to the centre of garment of interest. As a result, at the end of first step all the customers are mapped to at most one garment of interest. Also, every cluster will has exactly one garment of interest in it (and the tracking id of that garment of interest is used to identify the cluster i.e., \textit{Cluster-ID} equals \textit{Tracking-ID} of garment of interest).

In the next step, a membership table (in the form of a matrix) is generated which stores the information of clusters (represented by their respective \textit{centroid}) and their surrounding \textit{point}'s. A given point has a membership of \textit{1} if it is present in a cluster, else it is \textit{0}. Using \textit{maxDist} as a threshold, any \textit{point} representing the coordinates of a customer can be assigned to another cluster, if its distance from the \textit{centroid} of that cluster is not more than \textit{maxDist}, by setting it's membership value for that cluster to \textit{1}. In this manner, clusters can have overlapping \textit{point}'s which indicate that a customer is paying attention to multiple garments of interest at the same time.

The extended MCOKE algorithm is initially applied to the customers and garments of interest information, obtained from the first frame of the surveillance video. Afterwards, it is only applied for those frames where there is a significant change in the location of \textit{point} / \textit{centroid} corresponding to the customer / garments of interest, respectively. The coordinates of the centre of the bounding box of a garment or customer when the extended MCOKE algorithm was last applied are treated as “original coordinates”. A significant change occurs when, the distance between the current centre of the bounding box of a customer/garment of interest from the original coordinates, exceeds a specified value (\textit{mindist}). A significant change also occurs if a new customer/garment of interest is detected (or) an existing one is removed in the given frame. In such a case, the $k$ value is set to the total number of garments of interest present in the frame, before recomputing the clusters with the extended MCOKE. After the re-computation of clusters, the \textit{Cluster-ID} is assigned to the \textit{Tracking-ID} of the garment of interest present in it. This will ensure that a cluster is always identified by the single garment of interest present within it.

By monitoring the clusters and their members, a customer's interactions with garments in the store is tracked. Additional information such as the amount of time a customer spent looking at a particular garment, the colours of the garments he finds interesting, his expressions while glancing at the garments of interest and the amount of time he spends before making a decision, are obtained by monitoring the \textit{point}'s present in a cluster. This analysis, when combined with customer information such as, his age-group and gender will provide insight into the preferences and traits of customers belonging to a particular age-group and gender. Hence, sales analytics based on customer demographics, time spent on each garment, the garments of interest and the expressions during shopping may provide vital insight into the preferences of the customer base.

\section{Experimental Evaluation}\label{study}
In this section, we elaborate the experimental set-up and the dataset used to evaluate the proposed approach.  The video dataset used in this work is obtained from a retail garment store with two surveillance cameras: Stat Vision KC-ID2040D \cite{refx2} capturing the front-view, and Keeper CCD Video Camera KC-D3142 \cite{refx1} capturing the side view. This surveillance video is processed in three phases, with [\textit{phase-1}] to extract garment information,  [\textit{phase-2}] to capture customer related information like demography, facial expression etc., and the final [\textit{phase-3}] to utilize the information obtained in the earlier phases to conduct sales analytics.  The experiments were conducted on Google Colab \cite{refx3} in an environment with Intel(R) Xeon(R) 2.00 GHz CPU, NVIDIA Tesla T4 GPU, 16 GB GDDR6 VRAM and 13 GB RAM. All programming environment is Python 3.6 and OpenCV 4.2.

\subsection{Dataset}
\begin{figure*}
    \centering
    \includegraphics[width=160mm, height=210mm]{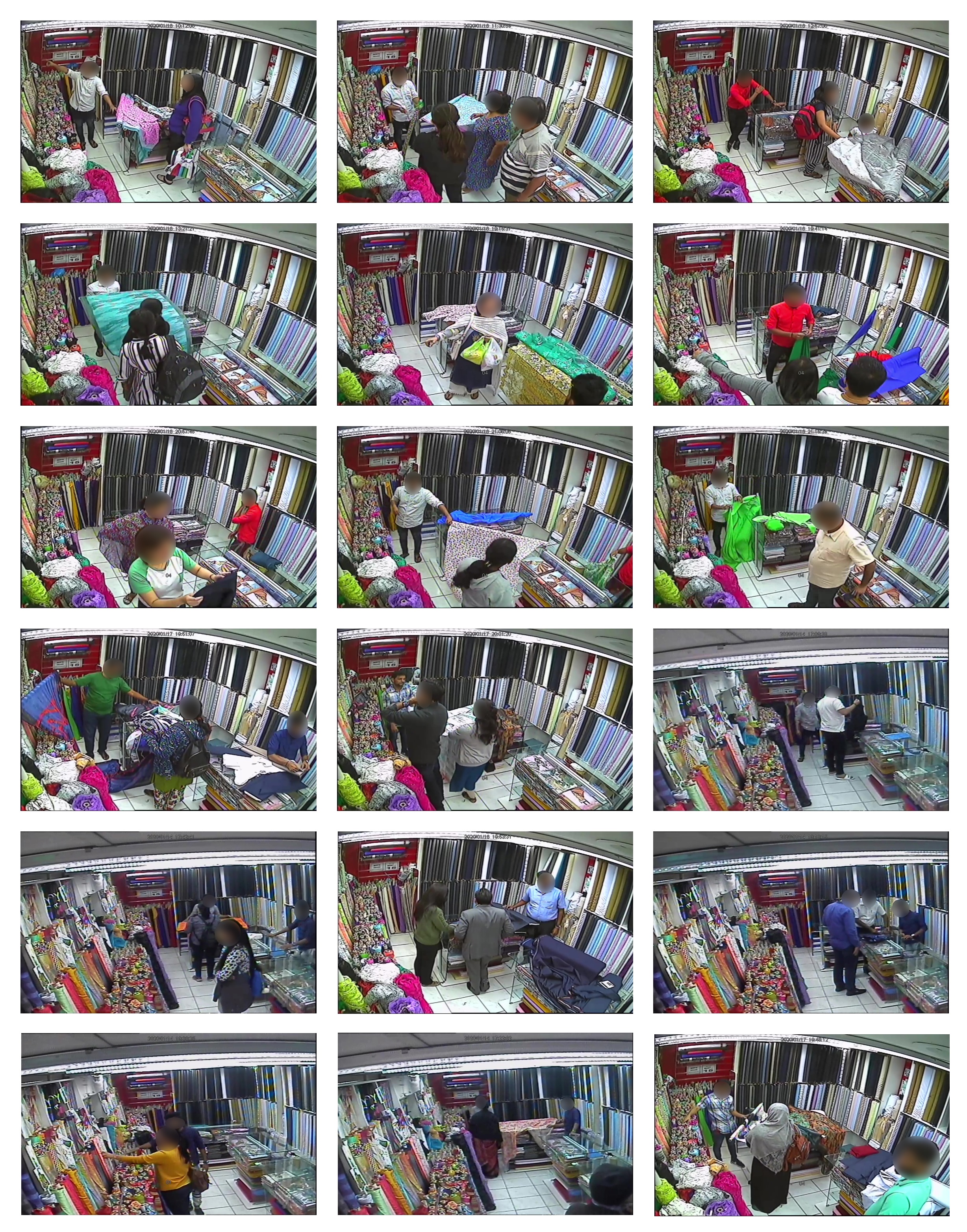}
    \caption{Some of the sales videos in the dataset}
    \label{fig:dataset}
\end{figure*}

The objective of the study is to analyse the behaviour of customers in real-world retail environment through CCTV footage, which is a challenging task due to the typically low resolution cameras, poor illumination conditions and environmental noise in real-life garment stores. The resolution of the videos is 944 $\times$ 576p and Figure \ref{fig:dataset} depicts some of the video samples in this dataset. The raw surveillance video in the dataset is pre-processed to obtain the customer and garment information, which is used in this work for sales analytics. In pre-processing, to identify the contours of the garments in [\textit{phase-1}], a Mixture of Gaussians \cite{ref5} model is used to identify the foreground, that is analysed using computer vision techniques such as image segmentation to detect the various garments of interest. To extract customer information in [\textit{phase-2}], a Wide Resnet WRN $16-8$ \cite{ref21} is used to predict the age and gender demographics; while mini Xception \cite{ref19}  model is used to obtain facial expressions of the customers. 

\subsection{Experimental results and statistics}
This section presents the results obtained by evaluating the proposed approach on the retail sales videos in this dataset. The sales information is analysed with respect to the gender of the customer, the age-group of the customer for each gender. The amount of time spent by customers during shopping for both genders is also analysed. The colour preference of various ages across both genders is studied with respect to the expressions exhibited by the customer when interacting with the garments. The customer's age is divided into a limited set of age groups \{\textit{child}, \textit{youth}, \textit{middle-aged}, \textit{elderly}\} based on their common interests/preferences as given in Table \ref{tab:ageGrp}. Details of the customer demographics obtained are listed below.

\begin{table}[h]
\begin{center}
\def\arraystretch{1.2}
\begin{tabular}{| p{2cm} | c |} 
\hline
 Age Range & Age Group \\
\hline
  1 to 17 years & \textit{child} \\ 
 18 to 29 years & \textit{youth} \\
 30 to 49 years &  \textit{middle-aged} \\
 50 to 90 years & \textit{elderly} \\
\hline
\end{tabular}
\caption{Age groups and their corresponding age ranges}
\label{tab:ageGrp}
\end{center}
\vspace{-4mm}
\end{table}

\begin{figure}[h]
\def\arraystretch{1.5}
	\begin{tabular}{c}
	\includegraphics[width=80mm, height=50mm]{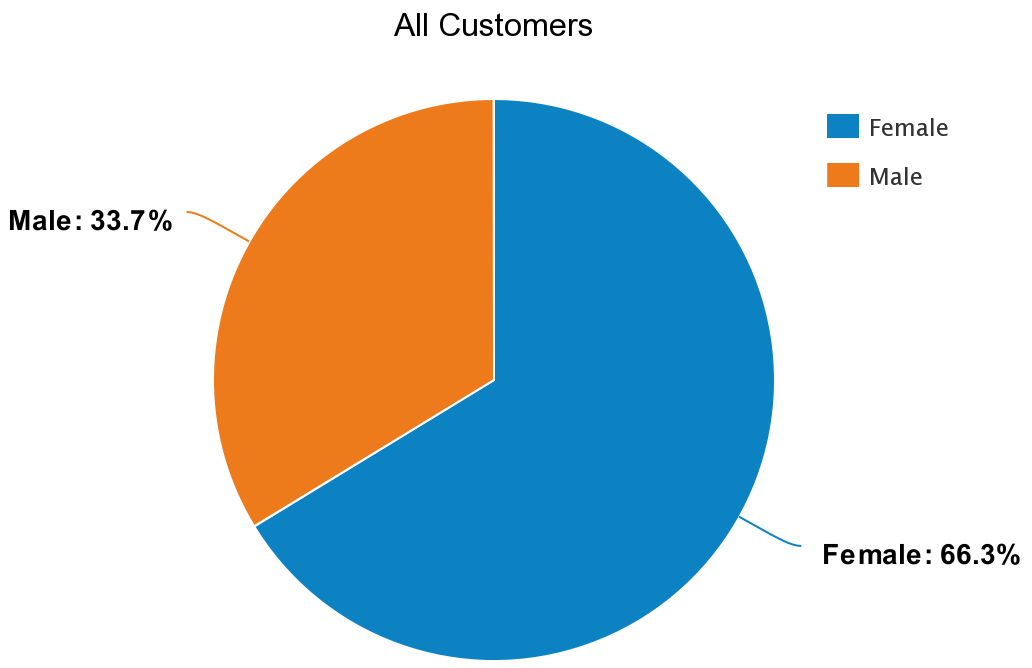}\\
	(a) Percentage of customers based on gender.\\
	\includegraphics[width=80mm, height=50mm]{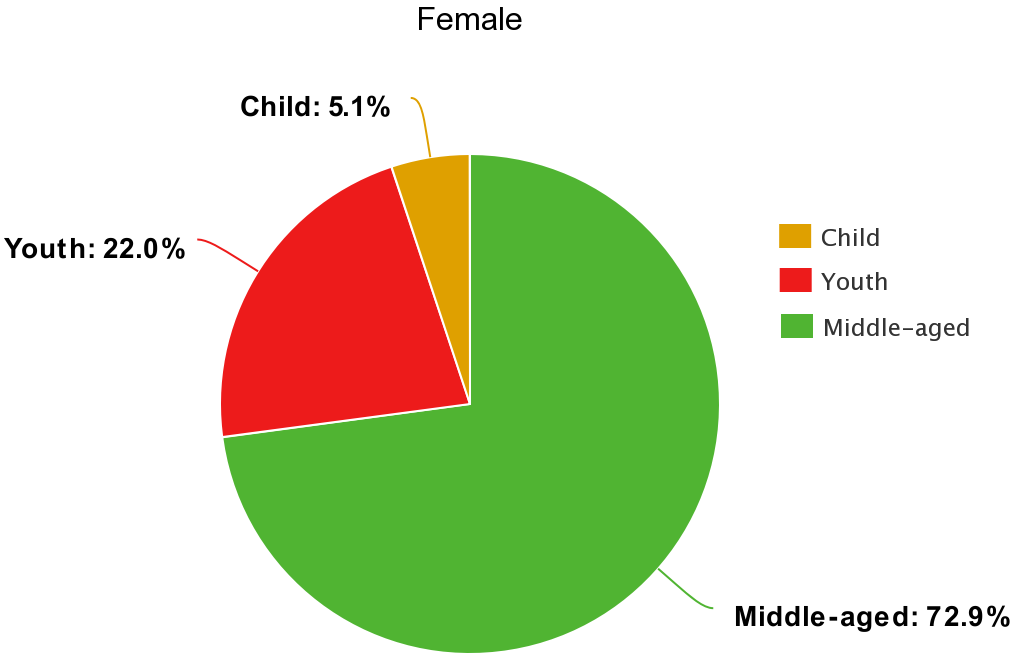}\\
	(b) Percentage of female customers based on age-group.\\
	\includegraphics[width=80mm, height=50mm]{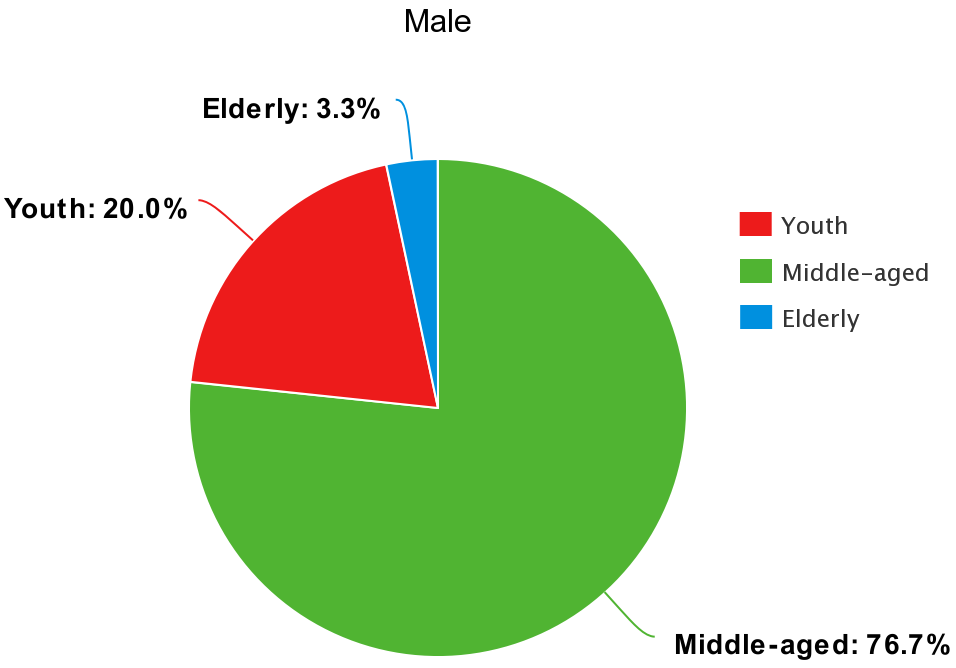}\\
	(c) Percentage of male customers based on age-group.
	\end{tabular}
	\caption{Customer's based on gender and age-group}
	\label{fig:population}
\vspace{-4mm}
\end{figure}

The ratio of customers based on gender, the percentage of female customers belonging to various age groups and  the percentage of male customers belonging to various age groups in the sales videos is depicted in Figures \ref{fig:population}(a), \ref{fig:population}(b) and \ref{fig:population}(c) respectively. The plots suggest that (i) there are more female customers than male customers, (ii) majority of the female and male customers are \textit{middle-aged}.

\begin{figure}[h]
    \centering
    \includegraphics[width=90mm, height=52mm]{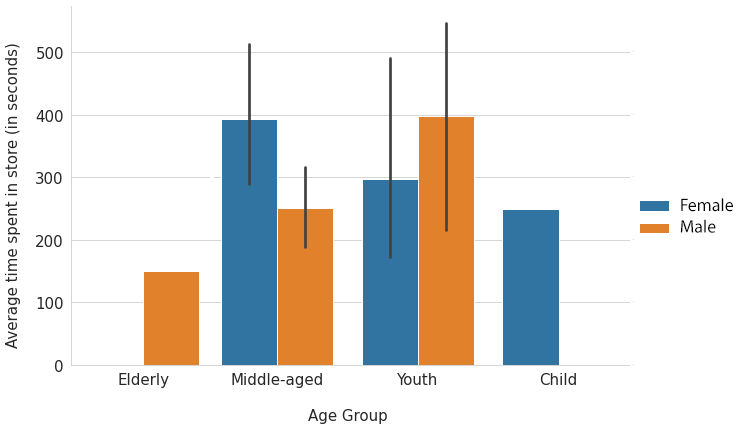}
    \caption{Average time spent in store}
    \label{fig:avgtime}
\vspace{-4mm}
\end{figure}

The average time spent by a customer of a specific \{age, gender\} pair, in shopping is shown through a candle plot in Figure \ref{fig:avgtime}. The observations from the figure are: (i) middle-aged women spend more time than men on shopping, (ii) young men spend more time than young women in shopping.

\begin{figure}
\def\arraystretch{1.5}
    	\begin{tabular}{c}
    	\includegraphics[width=90mm, height=50mm]{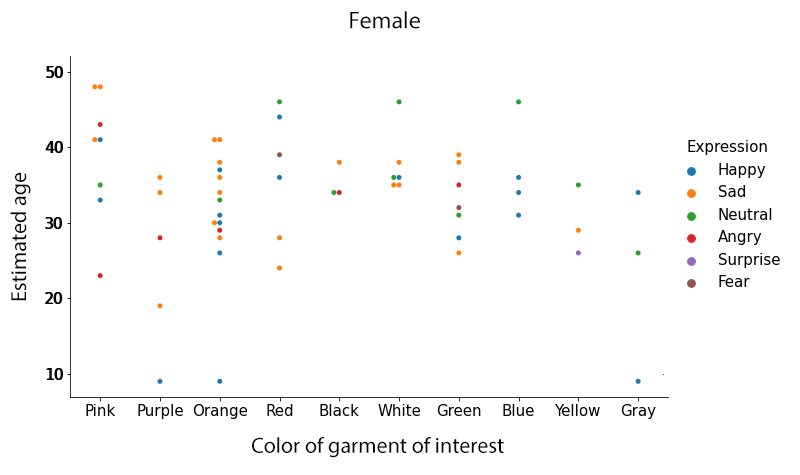}\\
	(a) Expressions of female customers by age against colour. \\
	\includegraphics[width=90mm, height=50mm]{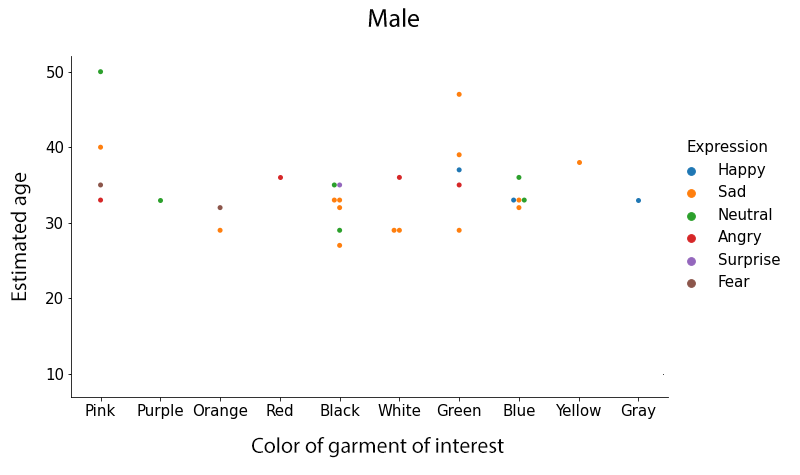}\\
	(b)  Expressions of male customers by age against colour.
	\end{tabular}
    \caption{Expression exhibited by customers, for a particular colour of garments of interest}
    \label{fig:emotion}
\vspace{-4mm}
\end{figure}

A scatter plot of the expressions exhibited by the customers for different garment colours is computed. The age is plotted along the \textit{y-axis}. The colour of the garment of interest is plotted along \textit{x-axis}. Separate scatter plots for female and male customers are shown in Figure \ref{fig:emotion}(a) and Figure \ref{fig:emotion}(b) respectively. From these plots, it can be observed that (i) women were more excited to look at various shades of Orange colour garments than men, and (ii) men preferred garments with \{Green, Blue, Gray\} colour.

\begin{figure}
        \begin{tabular}{c}
    	\includegraphics[width=80mm, height=60mm]{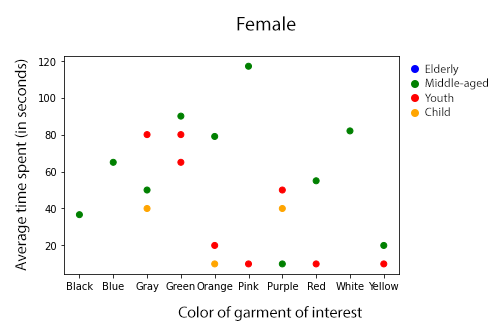}\\
	(a) Time spent by female customers \\on garments of various colours\\
	\includegraphics[width=80mm, height=60mm]{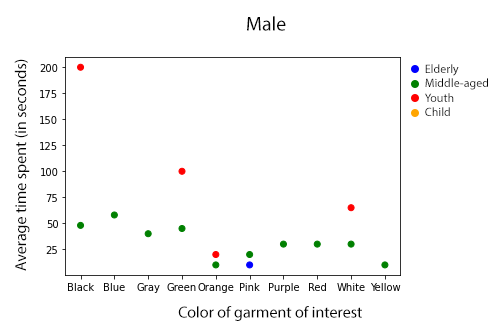}\\
	(b) Time spent by male customers \\on garments of various colours\\
	\end{tabular}
    \caption{Time spent by customer in observing garments of various colours.}
    \label{fig:colortime}
\vspace{-4mm}
\end{figure}

The average time spent by Female and Male customers against the colour of the garment of interest is depicted in Figures \ref{fig:colortime}(a) and \ref{fig:colortime}(b) respectively. From the figures, it can be concluded that (i) \textit{middle-aged} women spent more time on Pink, Green, White and Orange garments and (ii) young men spent more time on Blue, Green and White clothes.

The key observations from this study on this garment store are:
 \begin{enumerate}
   \item There were more female customers than male customers in this store.
   \item The \textit{middle-aged} customers are more than other ages.
   \item The \textit{middle-aged} female customers spend more time than male customers, looking at the various aspects of the garments.
   \item Female customers showed more interest towards garments of various shades of Orange colour.
 \end{enumerate}

In this approach, the customer is associated with a garment based on his/her position with respect to the garment and without considering any behavioral traits such as head orientation. Hence, there is the possibility that a customer may be falsely associated to a garment which he is not interested in, as it is close to him.

\section{Conclusions and Future Work}\label{conc}
In this work, a new approach is proposed to analyse sales videos to identify the garments of interest to a customer. This approach when applied to multiple sales videos will reveal statistics of the customer base and their preferences. This work relies on the prior computation of garments and customers information in sales videos, which is used in associating customers with garments they find interesting. It also obtains insightful data on the customer population of the store, and the garments they find interesting, along with the parameters such as a customer's expression and the time spent observing a garment. These results will help predict demand of different garments for various customer demographics. 

In this approach, a customer is placed in a cluster associated with a garment of interest based on his distance from the garment only. The future work can consider other behavioural traits of customers like orientation of the head to identify his garments of interest.

\section{Acknowledgement}
This work was done by Mr. Aniruddha Srinivas Joshi, Mr. Goutham Kanahasabai and Ms. Keerthi Priyanka; the final year B.Tech. students under the guidance of Dr. Earnest Paul Ijjina (Assistant Professor), in Department of Computer Science and Engineering, National Institute of Technology Warangal, as a part of their final year dissertation. The authors express their gratitude to the CSE department and the Institute, NIT Warangal for their efforts in developing the research environment, where this study was conducted. We also thank Google Colab for providing access to computational resources used to conduct this study.

\bibliographystyle{IEEEtran}
\bibliography{main}
\end{document}